\documentclass[epsfig,preprint,aps]{revtex4}
\usepackage{epsfig}
\begin{document}
\title{$q$-DEFORMED KINK SOLUTIONS}
\author{A. F. de
Lima$^{(a)}$ and R. de Lima Rodrigues$^{(b)*}$\thanks{Permanent
address: Departamento de Ci\^encias Exatas e da Natureza,
Universidade Federal de
Campina Grande, Cajazeiras, PB -- 58.900-000 -- Brazil.}\\
{}$^{(a)}$Departamento de F\'\i sica, Universidade Federal de Campina
Grande \\
 Campina Grande, PB -58.109-970 -- Brazil\\
{}$^{(b)}$ Centro Brasileiro de Pesquisas F\'\i sicas, Rua Dr. Xavier
Sigaud, 150\\ Rio de Janeiro-RJ-22290-180, Brazil}


\begin{abstract}
The $q$-deformed kink of the $\lambda\phi^4-$model is obtained via the
normalisable ground state eigenfunction of a fluctuation operator
associated with
the $q$-deformed hyperbolic functions. From such a bosonic zero-mode
the $q$-deformed potential in 1+1 dimensions is found, and we show that
the $q$-deformed kink solution
 is a kink displaced away from the origin.

\vspace{1cm}




\vspace{5cm} E-mail to RLR rafaelr@cbpf.br or
rafael@df.ufcg.edu.br, and to AFL aerlima@df.ufcg.edu.br.


To appear in International Journal of Modern Physics A [Particles
and Fields; Gravitation; Cosmology; Nuclear Physics], Vol. 21, No.
17 (2006) 3605-3613.

\vspace{1cm} {\small *Permanent address: Unidade Acad\^emica de
Ci\^encias Exatas e da Natureza, Universidade Federal de Campina
Grande, Cajazeiras, PB, 58.900-000, Brazil.}
\end{abstract}

\maketitle

\newpage

\section{introduction}

The kink of a scalar potential in 1+1 dimensions is a static,
non-singular, {\it classically stable} and a finite localized energy
solution of the equation of motion, which can be in topologically stable
sectors \cite{raja}.
In a recent lecture \cite{vacha02},
an investigation on the topological defects starting with the simplest case
of domain walls was presented, and then considerations to more
elaborate and realistic models were put forward.

In the present letter, one works
with the algebraic technique of the supersymmetry in
quantum mechanics (SUSY QM)  formulated by Witten
\cite{Witten81,La,Fred}, which is associated with a second order
differential equation for the $q$-deformed hyperbolic functions given
in Ref. \cite{arai}.
Recently, the $q$-deformed hyperbolic function was used to construct a new
$\eta-$pseudo-Hermitian
 complex potential with PT symmetry \cite{sun02}. Other potentials
like Rosen-Morse well, Scarf, Eckart and the generalized P\"oschl-Teller
were constructed  via shape invariance \cite{jia02}.

The stability equation for topological and non-topological solitons has been
approached in the framework of SUSY QM
\cite{Talu,Kuls,Sukumar86,K87,R95,vvr02}.

In this letter, the interesting program of proposing a new potential model in
1+1 dimensions, whose essential point is associated with the translational
invariance of the $q$-deformed kink solutions, is investigated.
We show that using the $q$-deformed hyperbolic functions which were
introduce by Arai \cite{arai}, the $q$-deformed kink solution,
is actually the known kink,
displaced away from
the origin of the $x$-axis.

\section{Solitons in 1+1 dimensions}

Consider the Lagrangian density for a single scalar field, $\phi(x,t),$ in
(1+1)-dimensions, in natural system, given by

\begin{equation}
{\cal L}\left(\phi, \partial_{\mu}\phi\right) = \frac{1}{2}
\partial_{\mu}\phi\partial^{\mu}\phi - V\left(\phi\right)
\label{E1},
\end{equation}
where  $V(\phi)$ is any positive semi-definite function of $\phi.$
It represents a
well-behaved potential energy. However, as it will be shown below, we have
found a new potential which is exactly solvable in the context of
the classical theory in (1+1)-dimensions.

The field equation for a static classical configuration, $\phi =
\phi_c\left(x\right),$ becomes

\begin{equation}
\label{E2}
-\frac{d^2}{dx^2} \phi_c\left(x\right) +
\frac{d}{d\phi_c}V\left(\phi_c\right) =0, \qquad \dot\phi_c = 0,
\end{equation}
with the following boundary conditions: $\phi_c(x)\rightarrow
\phi_{vacuum}(x)$ as $x\rightarrow \pm \infty.$

Since the potential is positive, it can be written as

\begin{equation}
\label{E4} V(\phi) = \frac{1}{2} U^2(\phi).
\end{equation}
Thus, the total energy for the $q$-kink becomes

\begin{eqnarray}
\label{ET}
E&&=\frac 12\int_{-\infty}^{\infty}\left[\left(\phi^{\prime}\right)^2+U^2\right]dx\nonumber\\
{}&&=\frac 12\int_{-\infty}^{\infty}\left[\left(\phi^{\prime}\mp U\right)^2\pm
2U\phi^{\prime}\right]dx.
\end{eqnarray}
In this case, the  Bogomol'nyi form of the energy, consisting of a
sum of squares and surface terms, becomes

\begin{equation}
\label{Ebogo}
E\geq\left|\int_{-\infty}^{\infty} dx\frac{\partial}{\partial x}U[\phi(x)]\right|=W_{12},
\end{equation}
under the well-known Bogomol'nyi condition for the kink solution,

\begin{equation}
\label{E5}
\frac{d\phi}{dx} = \pm U(\phi)
\end{equation}
where the solutions with the plus and minus signs represent two static
field configurations, and the superpotential $W(\phi)$ in field theory
must satisfy the following condition:

\begin{equation}
\label{ESP}
\frac{\partial}{\partial\phi}W(\phi) = U(\phi).
\end{equation}
The Bogomol'nyi mass bound of the energy result in a topological charge it is
given by
$W_{12}=W[\phi_2]-W[\phi_1],$ where $\phi_1$ and $\phi_2$
represent the vacuum states.

\section{Stability Equation}

The classical stability of the soliton solution is investigated by
considering small perturbations around it,

\begin{equation}
\label{E11} \phi(x,t) = \phi_c(x) + \eta (x,t),
\end{equation}
where we expand the fluctuations in terms of the normal modes,
\begin{equation}
\label{E12} \eta (x,t) = \sum_n \epsilon_n \eta_n (x) e^{i\omega_n t},
\end{equation}
with the $\epsilon_n'$s chosen so that $\eta(x,t)$ is real. A localized
classical configuration is said to be dynamically stable if the fluctuation
does not destroy it. The equation of motion becomes a Schr\"odinger-like
equation, viz.,

\begin{equation}
\label{E13}
O_F\eta_n (x) = \omega_n{^2}\eta_n (x), \quad
O_F=-\frac{d^2}{dx^2}+ V^{\prime \prime}(\phi_c),
\end{equation}
where $O_F$ is the fluctuation operator.
According to (\ref{E4}), one  obtains the supersymmetric form \cite{K87,R95}

\begin{equation}
\label{E14}
V^{\prime\prime}(\phi_c) = {U^\prime}^2 (\phi_c) +
U(\phi_c)U^{\prime \prime}(\phi_c),
\end{equation}
where the primes stand for a second derivative with respect to the
argument.

If the normal modes of (\ref{E13}) satisfy $\omega_n{^2} \geq 0,$  the
stability of the Schr\"odinger-like equation is ensured. Now, we are able
to implement a method that provides a new potential from  the potential
term that appears in the fluctuation operator.

Next, we consider the following generalized potential as corresponding to
the potential part of the fluctuation operator:

\begin{equation}
V^{\prime \prime}(\phi_c)=V(x;q) =
m^2\left[2-3qsech^2_q\left(\frac{m}{\sqrt 2}x \right)\right],
\end{equation}
where $q>0$ and we are using the $q$-deformed hyperbolic functions which were
introduce by Arai \cite{arai}:

\begin{eqnarray}
\label{qHF}
cosh_q(x)&&=\frac{e^x+qe^{-x}}{2}\nonumber\\
sinh_q(x)&&=\frac{e^x-qe^{-x}}{2}\nonumber\\
tanh_q(x)&&=\frac{sinh_q(x)}{cosh_q(x)}\nonumber\\
sech_q(x)&&=\frac{1}{cosh_q(x)}
\end{eqnarray}
where $x\epsilon${\bf R}. Thus

\begin{eqnarray}
\label{qHFd}
\frac{d}{dx}cosh_q(x)&&=sinh_q(x)\nonumber\\
\frac{d}{dx}sinh_q(x)&&=cosh_q(x)\nonumber\\ \frac{d}{dx}tanh_q(x)&&=
qsech_q^2(x)\nonumber\\
 \frac{d}{dx}sech_q(x)&&=
-tanh_q(x)sech_q(x)\nonumber\\
tanh_q^2(x)&&+qsech_q^2(x)=1.
\end{eqnarray}

The $q$-deformed potential term provides a fluctuation operator, so that their
eigenvalues satisfy the condition $\omega_n{^2}
\geq 0,$ and the ground state associated to the zero mode ($\omega_0^2=0$)
is given by

\begin{equation}
\label{GS}
\eta^{(0)} (x;q) =Nsech^{2}_q\left(\frac{m}{\sqrt 2}x \right),
\end{equation}
where $N$ is the normalization constant. Thus, the stability of the
Schr\"odinger-like equation is ensured.

The potential model we are now going to study presents translational
invariance, then, the bosonic zero-mode eigenfunction of the stability
equation is related with the kink by

\begin{equation}
\label{MZ}
\phi_q(x)= \int^{x}\eta^{(0)} (y;q)dy,
\end{equation}
so that, a priori, we may find the static classical configuration by a
first integration. Therefore, the potential model

\begin{equation}
\label{qpot}
 V(\phi;q)=\frac 12\left(\frac{d}{dx}\phi(x;q)\right)^2
\end{equation}
yields a class of $q$-deformed scalar potentials, $V(\phi)=V(\phi;q),$
which have exact
solutions.

Expressing the position coordinate in terms of the kink, i.e.
$x=x(\phi_k),$ then, from (\ref{GS}) and (\ref{MZ}) we obtain the $q$-deformed
kink

\begin{equation}
\label{K}
\phi (x;q) = \frac{m}{\lambda q}tanh_q\left(\frac{m}{\sqrt 2}x
\right).
\end{equation}
The explicit form of the $q$-kink for few values of $q$ is
shown in Fig. 1.

From equations (\ref{ESP}), (\ref{qp}) and (\ref{K}), we find a $q$-deformed
$\phi^4-$potential model with spontaneously
broken symmetry in scalar field theory and the superpotential, respectielly, viz.,

\begin{equation}
\label{PD}
V(\phi;q) = \frac{\lambda^2q^2}{4}\left(\phi^2_q -
\frac{m^2}{\lambda^2q^2}\right)^2, \quad W(\phi;q) =
\frac{\lambda q}{\sqrt{2}}\left(\frac 13 \phi^3_q -\frac{m^2}{\lambda^2q^2}\phi_q\right).
\end{equation}
It represents a well-behaved potential energy. Note that the $q$-deformed
$\phi^4$ model has a discrete symmetry as $\phi_q\rightarrow-\phi_q$ but it is
spontaneously broken for the vacuum state by the existence of two
degenerate vacua:

\begin{equation}
\label{V}
\phi_{1}= \frac{m}{q\lambda}, \qquad \phi_{2}= -\frac{m}{q\lambda}.
\end{equation}

The fact that the energy is finite is ensured because the kink
by the behavior of the approaches
one of the vacuum solutions at $\pm\infty.$ In the $q$-deformed
$\phi^4$ model there are four topological sectors, which are represented
by two
spaces $\Gamma_1$ and $\Gamma_2$ containing the $q$-deformed vacuum
solutions $\phi_1$
and $\phi_2$ and two spaces $\Gamma_3$ and $\Gamma_4$ containing the kink
and the anti-kink solutions.

The energy density of the $q$-kink is given by

\begin{equation}
\label{DE}
\epsilon(x)=\frac 12 \left[\left(\phi^{\prime}\right)^2+U^2\right]=
\frac{m^4}{2\lambda^2}sech_q^4\left(\frac{m}{\sqrt 2}x \right).
\end{equation}

Therefore, the kink mass or the so-called classical mass of the
pseudoparticle is given by

\begin{equation}
\label{MK}
M_{cl} = \int_{-\infty}^{+\infty}\epsilon(x)dx
=\frac{\sqrt{2}}{3}\left(\frac{m^3}{\lambda^2q^2}\right)
\end{equation}
which is dependent of $q.$ Note that when $q=1$
the undeformed case is re-obtained. In figure 2, we plot
the energy density given by Eq.(\ref{DE}), for few values of $q$. Also, the
Bogomol'nyi mass bound of the energy becomes
$E_B=\mid W_{12}\mid=M_{cl}.$

The  conserved topological
current becomes:

\begin{equation}
\label{ct}
{\cal J}_{\mu}=\frac 12 \epsilon_{\mu\nu}\partial^{\nu}{\tilde\phi}_q, \quad
 {\tilde\phi}_q=\frac{q\lambda}{m}\phi_q, \quad
\partial^{\mu}{\cal J}_{\mu}=0,
\end{equation}
where $ \epsilon_{\mu\nu}$ is the antisymmetric tensor in two dimensions
$ \epsilon_{01}=- \epsilon_{10}=1$ and is zero for the case
with repeated index.
The kink number or conserved topological charge is given by

\begin{equation}
\label{carga}
Q=\int_{-\infty}^{+\infty}{\cal J}_{0}dx=\frac 12 [\lim_{x\rightarrow +\infty}
\tilde\phi_q(x)-
 \lim_{x\rightarrow -\infty}\tilde\phi_q(x)]=1,
\end{equation}
which does not generate symmetries of the Lagrangian density and, therefore,
$Q$ is not a Noether charge. However, this charge is absolutely conserved,
$\frac{d}{dt}Q=0,$ so that the $q-$kink represents stable particle-like states.
Thus, the $q$-kink states can not decay by quantum tunneling into the
vacuum.

From the $q$-deformed potential, one then obtains the supersymmetric form

\begin{equation}
\label{FS}
V_-(x;q)
= W_q^2(x) + W^{\prime}_q,
\end{equation}
where the prime mean a first derivative with respect to the argument,
and $W_q(x)= -U^{\prime}_q(\phi_k)$ is the $q$-deformed superpotential
associated to the $q$-kink solution.
Thus, the bosonic and fermionic sector fluctuation operators
are respectively given by
\begin{eqnarray}
\label{E 19}
O_{F-} &&= -\frac{d^2}{dx^2}+W^2_q+W^{\prime}_q\nonumber\\
O_{F+} &&= -\frac{d^2}{dx^2}+W^2_q-W^{\prime}_q,
\end{eqnarray}
where $W_q(x) = -\sqrt{2}mtanh_q\left(\frac{m}{\sqrt 2}x \right).$
These fluctuation operators  are also called the supersymmetric partners,
which are isospectral up to the ground state.
The shape invariance condition of the pair of SUSY partner
potential will be investigated in a forthcoming paper.

Now let us show that the q-deformed potential of the stability equation
may be obtained from a translation of the position coordinate $x$ by $x-a,$
for the following choosing of the parameter $q:$

\begin{equation}
\label{qp}
q=e^{2a}.
\end{equation}
Indeed, when one substitutes the Eq. (\ref{qp}) in Eq. (\ref{E 19}) for the
q-deformed supersymmetric partners we find

\begin{eqnarray}
\label{Ept}
V_{-}(x-a;q=1)\rightarrow V_{-}(x;q)&&=
m^2\left[2-3qsech^2_q\left(\frac{m}{\sqrt 2}x \right)\right]\nonumber\\
V_{+}(x-a;q=1)\rightarrow V_{+}(x;q)&&=
m^2\left[2-qsech^2_q\left(\frac{m}{\sqrt 2}x \right)\right],
\end{eqnarray}
non-deformed supersymmetric ones.

Therefore, the q-deformed potentials recently investigated in literature
can be investigated, for a particular choice given by Eq. (\ref{qp}) as
corresponding to a simple translation. In this case the energy eigenvalues
are the same under the identification of their potential parameters.
Inversely, we can find from a non-deformed hyperbolic potential a class
of q-deformed hyperbolic ones by a simple translation of the variable
$x\rightarrow x-a,$ with $a=\ell n\left(\sqrt q \right),$ so that the
$q-$kink and vacuum states become
$
 \phi (x-a)) = \frac{m}{\lambda}e^{-2a}tanh\left(\frac{m}{\sqrt 2}(x-a)
\right)$ and $\pm\frac{m}{\lambda}e^{-2a},$ respectively.

\section{Conclusion}

In conclusion, we can say that, starting from a potential $V(x;q)$
in terms of the $q$-deformed
hyperbolic functions in the stability equation,
we construct the $q$-deformed topological kink associated
to the $q$-deformed $\phi^4$ potential model. We shown that the $q-$kink
mass is dependent of $q.$

We stress that a very rich spectrum of the states
(the $q-$kink and the quantum excitations about them), which was totally
unexpected in this model has emerged because of the existence of q-soliton
solutions.
However, what we have call
$q$-deformed kink solution, is actually the known kink, displaced away from
the origin of the $x$-axis. Specifically,
if we set $q=exp(2a)$ the $q$-deformed
hyperbolic tangent is just the ordinary $tanh$ with its argument shifted by
$a,$ i.e.  $tanh_q(x)=tanh(x-a).$ So, the $q$-deformed kink (\ref{K})
is just the
known kink of the potential (\ref{PD}), centered around $(\ell nq)/2.$
This can be
checked on the graphs of Fig.1. Of course, the asymptotic value is
$\frac{m}{q\lambda},$ which is also shown in Fig.1.

This particular choice of the $q-$deformation parameter means that we can
map a $q$-deformed system into one with a simple translation. It is worth
to call the attention to the fact
that this new translated system is also a q-deformed system by the initial
correspondence we have established. All that we have done was to shown that
it is possible to find a suitable choice of the parameter in order
to describe the $q-$deformed one terms of a pure translation.

Finally, it is important to pointed out that one can extend our
approach to 3+1 dimensions.
Indeed, the present work opens a
new route for future investigations on
domain walls \cite{vacha02} from $q-$deformation of potential model in terms
of coupled scalar fields. For instance, let us point out
that  our approach can be
applied from two \cite{RV96,guila02a} and three \cite{RPV95}
coupled scalar fields, where in both cases the
soliton solutions only depend
on $z$ but not on $x$ and $y.$

\vskip 1.0 true cm

\noindent{{\large\bf Acknowledgments}}

RLR wishes to thanks the staff of CBPF and UACEN-CFP-UFCG for
facilities. This research was supported in part by CNPq (Brazilian
Research Agency). RLR is grateful to Prof. J. A. Helayel-Neto and
Prof. M. A. Rego-Monteiro for fruitful discussions.
 The authors would also like to thank A. Arai and S.-C. Jia for the
kind interest in pointing out relevant
references on the subject of this paper.


\unitlength=1cm
\begin{figure}[tbp]
\centering
\begin{picture}(10,1)
\epsfig{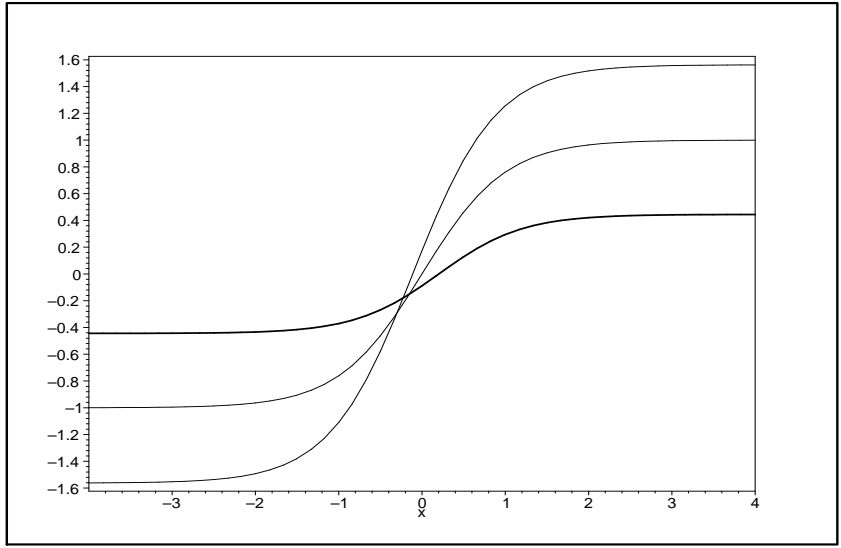}
\end{picture}
\vspace{7.5cm}
\caption{The $q$-deformed kink profile,
with $q=0.8$(thickness=1), $q=1.0$(dotted curve), and $q=3.0$(thickness=3),
respectively, for $m=\lambda=\sqrt{2}.$
}
\end{figure}


\unitlength=1cm
\begin{figure}[tbp]
\centering
\begin{picture}(10,1)
\epsfig{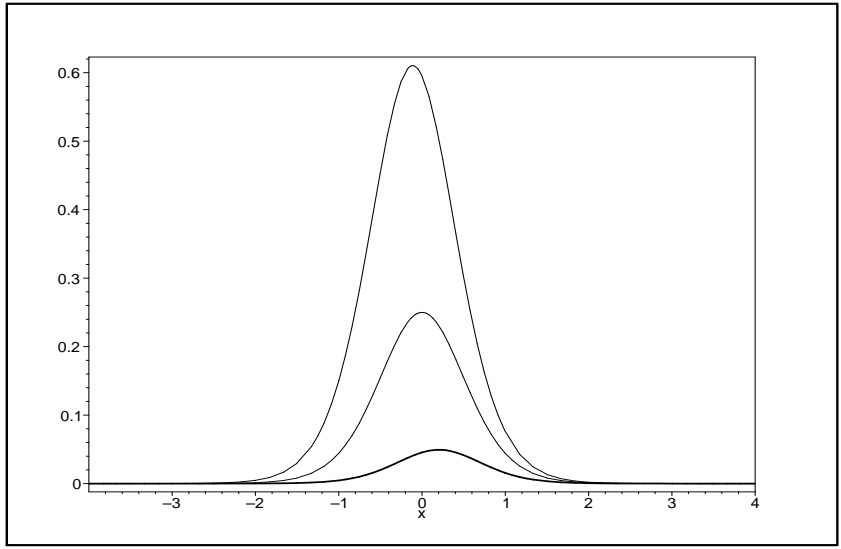}
\end{picture}
\vspace{7.5cm}
\caption{The energy density given by Eq.(\ref{DE}),
with $q= 0.8$(thickness=1), $q=1.0$(dotted curve), and $q=3.0$(thickness=3),
respectively, for $m=\lambda=\sqrt{2}.$}
\end{figure}

\end{document}